\documentclass[11pt]{article}
\usepackage{amssymb,amsmath,amsfonts}
\usepackage{graphicx}
\usepackage{graphics}
\usepackage{eepic,epsfig}

\begin{document}

\title{Casimir-Polder potential for a metallic cylinder \\
in cosmic string spacetime}
\author{A. A. Saharian\thanks{%
E-mail: saharian@ysu.am}, \thinspace\ A. S. Kotanjyan \\
\\
\textit{Department of Physics, Yerevan State University,}\\
\textit{1 Alex Manoogian Street, 0025 Yerevan, Armenia}}
\maketitle

\begin{abstract}
Casimir-Polder potential is investigated for a polarizable microparticle in
the geometry of a straight cosmic string with a metallic cylindrical shell.
The electromagnetic field Green tensor is evaluated on the imaginary
frequency axis. The expressions for the Casimir-Polder potential is derived
in the general case of anisotropic polarizability for the both interior and
exterior regions of the shell. The potential is decomposed into pure string
and shell-induced parts. The latter dominates for points near the shell,
whereas the pure string part is dominant near the string and at large
distances from the shell. For the isotropic case and in the region inside
the shell the both pure string and shell-induced parts in the Casimir-Polder
force are repulsive with respect to the string. In the exterior region the
shell-induced part of the force is directed toward the cylinder whereas the
pure string part remains repulsive with respect to the string. At large
distances from the shell the total force is repulsive.
\end{abstract}

\bigskip


\bigskip

\section{Introduction}

The production of cosmic strings in symmetry breaking phase transitions
during the evolution of the early universe is predicted by a wide class of
particle physics models \cite{Vile85}. The considerable attention attracted
by this class of topological defects was motivated by the fact that the
cosmic strings are candidates for the generation of a variety of interesting
physical effects. The latter include gravitational lensing, anisotropies in
the cosmic microwave background radiation, the generation of gravitational
waves, high-energy cosmic rays, and gamma ray bursts. More recently it has
been shown that cosmic strings form in brane inflation models as a by
product of the annihilation of the branes (for a review see \cite{Cope10}).

In quantum field theory, the non-trivial topology of space around a cosmic
string results in the distortion of the vacuum fluctuations of quantized
fields. This induces non-zero vacuum expectation values for physical
observables such as the field squared and the energy-momentum tensor. In a
previous paper \cite{Saha11EPJ}, we have shown that the distortion of the
vacuum fluctuations spectrum by the cosmic string also gives rise to
Casimir-Polder forces acting on a polarizable microparticle (see also \cite%
{Bard10} for the force in the static limit). These forces have attracted a
great deal of attention because of their important role in many areas of
science, including material sciences, physical chemistry, nanotechnology,
and atom optics (for reviews see \cite{Parseg}). In \cite{Saha11EPJ} it has
been shown that, in dependence on the eigenvalues for the polarizability
tensor and of the orientation of its principal axes, the Casimir-Polder
force induced by the string can be either repulsive or attractive. For an
isotropic polarizability tensor the force is always repulsive.

Another source for the vacuum polarization is the presence of material
boundaries. The boundary conditions imposed on a quantum field alter the
zero-point oscillations spectrum and lead to additional shifts in the vacuum
expectation values. Combined effects of topology and boundaries on the
quantum vacuum in the geometry of a cosmic string have been investigated
previously for scalar \cite{Beze06}, electromagnetic \cite{Brev95,Beze07}
and fermionic fields \cite{Beze08}, constrained on a cylindrical boundary
coaxial with the cosmic string. The analysis of the vacuum energy for
massless scalar fields subject to Dirichlet, Neumann and hybrid boundary
conditions in the setting of the conical piston has been recently developed
in \cite{Fucc11}. The vacuum polarization effects in a cosmic string
spacetime induced by a scalar field obeying Dirichlet or Neumann boundary
conditions on a surface orthogonal to the string are considered in \cite%
{Beze11}.

In the present paper we derive the exact Casimir-Polder (CP) potential for
the general case of frequency dependent anisotropic polarizability of a
microparticle in the geometry of straight cosmic string with a coaxial
conducting cylindrical shell. From the point of view of the physics in the
region outside the string, this geometry can be considered as a simplified
model for the nontrivial core. This model presents a framework in which the
influence of the finite core effects on physical processes in the vicinity
of the cosmic string can be investigated. The corresponding results may shed
light upon features of finite core effects in more realistic models. In
addition, the problem considered here is of interest as an example with
combined topological and boundary-induced quantum effects in which the CP
forces can be found in closed form. The CP interaction potential of a
microparticle with an ideal metal cylindrical shell in background of
Minkowski spacetime has been investigated in a number of papers, in
particular,  related to applications to carbon nanotubes (see, for instance,
\cite{Blag05}-\cite{Milt11} and references therein). Recently, the exact
potential for a microparticle outside a cylindrical shell has been found in
\cite{Eber09} using the Hamiltonian approach. The CP potential for both
regions inside and outside an ideal metal cylindrical shell is investigated
in \cite{Beze11b} using the Green function method. In this paper the exact
quantum field theoretical result is compared with that obtained using the
proximity force approximation and a very good agreement is demonstrated. In
\cite{Elli10} it was shown that for a particle out of thermal equilibrium
with its environment inside a cylindrical cavity the CP potential can be
enhanced by fine-tuning the cavity radius to resonate with the particle's
internal transition wavelength.

We have organized the paper as follows. In the next section we evaluate the
Green tensor in the frequency domain in the region inside a conducting
cylindrical shell in the geometry of a cosmic string. By using the
generalized Abel-Plana summation formula, the Green tensor is decomposed
into the boundary-free and boundary-induced parts. The corresponding CP
potential is investigated in section \ref{sec:CPin} for the general case of
anisotropic polarizability. The Green tensor and the CP potential for the
region outside a cylindrical shell are considered in section \ref%
{sec:GreenCPext}. Similar to the interior region, these quantities are
presented as the sum of pure string and shell-induced parts. Section \ref%
{sec:Conc} summarizes the main results of the paper.

\section{Electromagnetic field Green tensor inside a cylindrical shell}

\label{sec:GT}

In the cylindrical coordinates $(x^{1},x^{2},x^{3})=(r,\phi ,z)$, the
geometry of an idealized infinitely long straight cosmic string is described
by the line element
\begin{equation}
ds^{2}=dt^{2}-dr^{2}-r^{2}d\phi ^{2}-dz{}^{2},  \label{ds21}
\end{equation}%
where $0\leqslant r<\infty $, $-\infty <z<+\infty $, $0\leqslant \phi
\leqslant \phi _{0}$ and the spatial points $(r,\phi ,z)$ and $(r,\phi +\phi
_{0},z)$ are identified. The planar angle deficit is related to the mass $%
\mu _{0}$ per unit length of the string by $2\pi -\phi _{0}=8\pi G\mu _{0}$,
with $G$ being the Newton gravitational constant. (Effective metric with a
planar angle deficit also arises in a number of condensed matter systems
(see, for instance, \cite{Volo98}).) In addition, we shall assume the
presence of a coaxial metallic cylindrical shell of radius $a$.

The nontrivial topology due to the cosmic string and the boundary conditions
imposed for the electric and magnetic fields on the cylindrical shell change
the structure of the zero-point fluctuations of the electromagnetic field.
In particular, a neutral polarizable microparticle experiences a dispersion
force, the CP force. For a particle with the polarizability tensor $\alpha
_{jl}(\omega )$, the corresponding interaction potential is expressed in
terms of the subtracted Green tensor as (see \cite{Parseg})%
\begin{equation}
U(\mathbf{r})=\frac{1}{2\pi }\int_{0}^{\infty }d\xi \,\,\alpha _{jl}(i\xi
)G_{jl}^{\text{(s)}}(\mathbf{r},\mathbf{r};i\xi ),  \label{CPpot}
\end{equation}%
where $\mathbf{r}$ is the location of the microparticle and summation is
understood over the indices $j,l=1,2,3$. In (\ref{CPpot}),
\begin{equation}
G_{jl}^{\text{(s)}}(\mathbf{r},\mathbf{r}^{\prime };\omega )=\int_{-\infty
}^{+\infty }d\tau \,[G_{jl}(x,x^{\prime })-G_{jl}^{\text{(M)}}(x,x^{\prime
})]e^{i\omega \tau },  \label{Gjls}
\end{equation}%
where $G_{jl}(x,x^{\prime })$, with $x=(t,\mathbf{r})$, $x^{\prime
}=(t^{\prime },\mathbf{r}^{\prime })$, $\tau =t-t^{\prime }$, is the
retarded Green tensor for the electromagnetic field in the geometry of a
cosmic string with the cylindrical shell and $G_{jl}^{\text{(M)}%
}(x,x^{\prime })$ is the corresponding tensor in the boundary-free Minkowski
spacetime. The geometry of a cosmic string is flat outside the string core
and the renormalization procedure is reduced to the subtraction of the
Minkowskian part.

For the evaluation of the Green tensor in (\ref{Gjls}) we use the direct
mode summation method. Let $\{\mathbf{E}_{\alpha }(x),\mathbf{E}_{\alpha
}^{\ast }(x)\}$ be a complete set of normalized mode functions for the
electric field, specified by a collective index $\alpha $. For the Green
tensor we have the following mode sum formula:
\begin{equation}
G_{jl}(x,x^{\prime })=-i\theta (\tau )\sum_{\alpha }\left[ E_{\alpha
j}(x)E_{\alpha l}^{\ast }(x^{\prime })-E_{\alpha l}(x^{\prime })E_{\alpha
j}^{\ast }(x)\right] ,  \label{Gjlms}
\end{equation}%
where $\theta (\tau )$ is the unit-step function and the indices $j,l=1,2,3$
correspond to the coordinates $r,\phi ,z$, respectively.

First we consider the region inside the cylindrical shell. In the problem
under consideration we have two classes of mode functions corresponding to
the cylindrical waves of the transverse magnetic (TM, $\lambda =0$) and
transverse electric (TE, $\lambda =1$) types. The mode functions for the
electric field are obtained from the corresponding functions for the vector
potential given in Ref. \cite{Beze07} and they have the form%
\begin{equation}
\mathbf{E}_{\alpha }^{(\lambda )}(x)=\beta _{\alpha }\mathbf{E}^{(\lambda
)}(r)e^{iqm\phi +ikz-i\omega t},  \label{Ealf}
\end{equation}%
where $m=0,\pm 1,\pm 2,\ldots $, $-\infty <k<+\infty $, $\omega =\sqrt{%
\gamma ^{2}+k^{2}}$, and%
\begin{equation}
q=2\pi /\phi _{0}.  \label{qu}
\end{equation}
The radial functions $E_{l}^{(\lambda )}(r)$ in (\ref{Ealf}) are given by
the expressions
\begin{eqnarray}
E_{1}^{(0)}(r) &=&ik\gamma J_{q|m|}^{\prime }(\gamma r),\;E_{2}^{(0)}(r)=-%
\frac{kqm}{r}J_{q|m|}(\gamma r),\;E_{3}^{(0)}(r)=\gamma ^{2}J_{q|m|}(\gamma
r),  \notag \\
E_{1}^{(1)}(r) &=&-\frac{\omega qm}{r}J_{q|m|}(\gamma
r),\;E_{2}^{(1)}(r)=-i\omega \gamma J_{q|m|}^{\prime }(\gamma
r),\;E_{3}^{(1)}(r)=0,  \label{Eir}
\end{eqnarray}%
where $J_{\nu }(x)$ is the Bessel function, the prime means the derivative
with respect to the argument of the function. From the standard boundary
conditions for the electric and magnetic fields on the cylindrical boundary
with radius $a$, we can see that the eigenvalues for the quantum number $%
\gamma $ are roots of the equations
\begin{equation}
J_{q|m|}^{(\lambda )}(\gamma a)=0,\quad \lambda =0,1,  \label{modes1}
\end{equation}%
where $J_{\nu }^{(0)}(x)=J_{\nu }(x)$ and $J_{\nu }^{(1)}(x)=J_{\nu
}^{\prime }(x)$. In the discussion below the corresponding eigenmodes are
denoted by $j_{m,n}^{(\lambda )}=\gamma a$, $n=1,2,\ldots $. As a result the
set of quantum numbers specifying the eigenfunctions is given by $\alpha
=(k,m,\lambda ,n)$. The normalization coefficient in (\ref{Ealf}) is given
by the expression \cite{Beze07}
\begin{equation}
\beta _{\alpha }^{2}=\frac{qT_{q|m|}(\gamma a)}{\pi \omega a\gamma }%
,\;T_{\nu }(x)=\frac{x}{J_{\nu }^{^{\prime }2}(x)+(1-\nu ^{2}/x^{2})J_{\nu
}^{2}(x)}.  \label{betalf}
\end{equation}

Substituting the mode functions (\ref{Ealf}) into the mode sum formula (\ref%
{Gjlms}), the following representation is obtained for the Green tensor on
the imaginary frequency axis:%
\begin{eqnarray}
G_{jl}(\mathbf{r},\mathbf{r}^{\prime };i\xi ) &=&-\frac{q}{\pi }%
\,\sum_{m=-\infty }^{+\infty }\int_{-\infty }^{+\infty }dk\sum_{\lambda
=0,1}\sum_{n=1}^{\infty }\frac{T_{q|m|}(j_{m,n}^{(\lambda )})}{%
j_{m,n}^{(\lambda )}\omega _{m,n}^{(\lambda )}}  \notag \\
&&\times \left[ E_{j}^{(\lambda )}(r)E_{l}^{(\lambda )\ast }(r^{\prime })%
\frac{e^{iqm\Delta \phi +ik\Delta z}}{\omega _{m,n}^{(\lambda )}-i\xi }%
+E_{l}^{(\lambda )}(r^{\prime })E_{j}^{(\lambda )\ast }(r)\frac{%
e^{-iqm\Delta \phi -ik\Delta z}}{\omega _{m,n}^{(\lambda )}+i\xi }\right] .
\label{GjlC1}
\end{eqnarray}%
where $\Delta \phi =\phi -\phi ^{\prime }$ and $\Delta z=z-z^{\prime }$, $%
\omega _{m,n}^{(\lambda )}=\sqrt{j_{m,n}^{(\lambda )2}/a^{2}+k^{2}}$. For
the summation of the series over $n$ we apply the formula \cite{SahaBook}%
\begin{equation}
\sum_{n=1}^{\infty }T_{q|m|}(j_{m,n}^{(\lambda )})f(j_{m,n}^{(\lambda )})=%
\frac{1}{2}\int_{0}^{\infty }dx\,f(x)-\frac{\pi i}{2}\sum_{p}\underset{%
z=iy_{p}}{\mathrm{Res}}f(z)\frac{H_{q|m|}^{(1,\lambda )}(z)}{%
J_{q|m|}^{(\lambda )}(z)},  \label{Sumf}
\end{equation}%
where $z=\pm iy_{p}$, $y_{p}>0$, $p=1,2,\ldots $, are poles of the function $%
f(z)$ and $H_{\nu }^{(s,0)}(z)=H_{\nu }^{(s)}(z)$, $H_{\nu
}^{(s,1)}(z)=H_{\nu }^{(s)\prime }(z)$, with $H_{\nu }^{(s)}(z)$, $s=1,2$,
being the Hankel functions. In (\ref{Sumf}), it is assumed that $f(z)$ is an
analytic function for $\mathrm{Re\,}z>0$ and obeys the conditions $f(ye^{\pi
i/2})=-e^{2q|m|\pi i}f(ye^{-\pi i/2})$ and $|f(x+iy)|<\varepsilon (x)e^{by}$%
, $b<2$, for $y\rightarrow \infty $, with $\varepsilon (x)\rightarrow 0$ for
$x\rightarrow \infty $. For the poles of the function $f(z)$, corresponding
to the series in (\ref{GjlC1}), one has $y_{p}=a\sqrt{k^{2}+\xi ^{2}}$. The
part of the Green tensor corresponding to the first term on the right-hand
side of (\ref{Sumf}) is the Green tensor in the boundary-free cosmic string
geometry. The latter will be denoted by $G_{jl}^{(0)}(\mathbf{r},\mathbf{r}%
^{\prime };i\xi )$. The boundary-induced part of the Green tensor
corresponds to the second term.

As a result, by using (\ref{Sumf}), the Green tensor is presented in the
decomposed form%
\begin{equation}
G_{jl}(\mathbf{r},\mathbf{r}^{\prime };i\xi )=G_{jl}^{(0)}(\mathbf{r},%
\mathbf{r}^{\prime };i\xi )+G_{jl}^{\text{(b)}}(\mathbf{r},\mathbf{r}%
^{\prime };i\xi ).  \label{Gjldec}
\end{equation}%
The boundary-induced part is given by the formula
\begin{eqnarray}
G_{jl}^{\text{(b)}}(\mathbf{r},\mathbf{r}^{\prime };i\xi ) &=&-\frac{q}{\pi }%
\,\sum_{m=-\infty }^{\infty }e^{iqm\Delta \phi }\sum_{\lambda =0,1}(-\xi
^{2})^{\lambda }\int_{-\infty }^{\infty }dk\,e^{ik\Delta z}  \notag \\
&&\times k^{2(1-\lambda )}\frac{K_{q|m|}^{(\lambda )}(a\gamma )}{%
I_{q|m|}^{(\lambda )}(a\gamma )}i_{j}^{(\lambda )}(\gamma r,\gamma
/k)i_{l}^{(\lambda )\ast }(\gamma r^{\prime },\gamma /k),  \label{Gjl3}
\end{eqnarray}%
where in the integrand $\gamma =\sqrt{k^{2}+\xi ^{2}}$. In the last
expression, $I_{\nu }(x)$ and $K_{\nu }(x)$ are the modified Bessel
functions, $F_{\nu }^{(0)}(x)=F_{\nu }(x)$, $F_{\nu }^{(1)}(x)=F_{\nu
}^{\prime }(x)$ for $F=I,K$, and the functions $i_{l}^{(\lambda )}(x,y)$\
are defined as%
\begin{eqnarray}
i_{1}^{(0)}(x,y) &=&I_{q|m|}^{\prime }(x),i_{2}^{(0)}(x,y)=i\frac{qm}{x}%
I_{q|m|}(x),\;i_{3}^{(0)}(x,y)=iyI_{q|m|}(x),  \notag \\
i_{1}^{(1)}(x,y) &=&\frac{qm}{x}I_{q|m|}(x),\;i_{2}^{(1)}(x,y)=-iI_{q|m|}^{%
\prime }(x),\;i_{3}^{(1)}(x,y)=0.  \label{Etilde}
\end{eqnarray}%
In accordance with (\ref{CPpot}), for the evaluation of the CP potential we
need the expression of the boundary-induced part in the coincidence limit.
In this limit the off-diagonal components vanish and for the diagonal
components we \ have%
\begin{equation}
G_{ll}^{\text{(b)}}(\mathbf{r},\mathbf{r};i\xi )=-\frac{4q}{\pi }\,%
\sideset{}{'}{\sum}_{m=0}^{\infty }\sum_{\lambda =0,1}(-\xi ^{2})^{\lambda
}\int_{\xi }^{\infty }d\gamma \,\gamma \,\frac{K_{qm}^{(\lambda )}(a\gamma )%
}{I_{qm}^{(\lambda )}(a\gamma )}\frac{|i_{l}^{(\lambda )}(\gamma r,\gamma /%
\sqrt{\gamma ^{2}-\xi ^{2}})|^{2}}{(\gamma ^{2}-\xi ^{2})^{\lambda -1/2}}{},
\label{Gll}
\end{equation}%
where the prime on the summation sign means that the term $m=0$ should be
taken with the coefficient 1/2.

\section{Casimir-Polder potential}

\label{sec:CPin}

On the base of Eq. (\ref{Gjldec}), the CP potential in the presence of the
cylindrical shell is decomposed as:%
\begin{equation}
U(r)=U_{0}(r)+U_{\text{b}}(r),  \label{Udec}
\end{equation}%
where $U_{0}(r)$ is the potential for the geometry of a cosmic string
without boundaries and the term $U_{\text{b}}(r)$ is due the presence of the
cylindrical shell. The pure string part is investigated in \cite{Saha11EPJ}
and here we will be mainly concerned with the boundary-induced part. The
latter is given by the expression%
\begin{equation}
U_{\text{b}}(r)=\frac{1}{2\pi }\int_{0}^{\infty }d\xi \,\,\alpha _{jl}(i\xi
)G_{jl}^{\text{(b)}}(\mathbf{r},\mathbf{r};i\xi ).  \label{Ub}
\end{equation}%
As it has been mentioned before, the off-diagonal components of the boundary
induced part of the Green tensor in (\ref{Ub}) vanish. As a result, by
taking into account the expression (\ref{Gll}), we get%
\begin{eqnarray}
U_{\text{b}}(r) &=&-\frac{2q}{\pi ^{2}}\,\sideset{}{'}{\sum}_{m=0}^{\infty
}\sum_{\lambda =0,1}\sum_{l=1}^{3}\int_{0}^{\infty }d\xi \,\,\alpha
_{ll}(i\xi )(-\xi ^{2})^{\lambda }  \notag \\
&&\times \,\int_{\xi }^{\infty }d\gamma \,\gamma \frac{K_{qm}^{(\lambda
)}(a\gamma )}{I_{qm}^{(\lambda )}(a\gamma )}\frac{|i_{l}^{(\lambda )}(\gamma
r,\gamma /\sqrt{\gamma ^{2}-\xi ^{2}})|^{2}}{(\gamma ^{2}-\xi ^{2})^{\lambda
-1/2}}.  \label{Ubin}
\end{eqnarray}%
In the special case $q=1$ this formula reduces to the result of \cite%
{Beze11b} for the CP interaction potential for a cylindrical shell in
Minkowski spacetime.

In (\ref{Ubin}), $\alpha _{ll}(i\xi )$ are the physical components of the
polarizability tensor in the cylindrical coordinates corresponding to line
element (\ref{ds21}). These components depend on the orientation of the
polarizability tensor principal axes. As a consequence, the CP potential
depends on the distance of the microparticle from the string and on the
angles determining the orientation of the principal axes. Let us introduce
Cartesian coordinates $x^{\prime \prime l}=(x^{\prime \prime },y^{\prime
\prime },z^{\prime \prime })$ with the $z^{\prime \prime }$-axis along the
string and with the particle location at $(r,0,0)$ and let $\beta _{ln}$ be
the cosine of the angle between $x^{\prime \prime l}$ and the $n$-th
principal axis of the polarizability tensor. One has $\sum_{n=1}^{3}\beta
_{ln}^{2}=1$. Now we can write $\alpha _{ll}(\omega )=\sum_{n=1}^{3}\beta
_{ln}^{2}\alpha _{n}(\omega )$, where $\alpha _{n}(\omega )$ are the
principal values of the polarizability tensor. The coefficients $\beta _{ln}$
can be expressed in terms of the Euler angles determining the orientation of
the principal axes with respect to the coordinate system $x^{\prime \prime l}
$. In the isotropic case $\alpha _{n}(\omega )\equiv \alpha (\omega )$ and
we have $\alpha _{ll}(\omega )=\alpha (\omega )$.

The boundary-induced part of the potential is finite on the string. Assuming
that $q>1$, we can see that only the $m=0$ term contributes and%
\begin{equation}
U_{\text{b}}(0)=-\frac{q}{\pi ^{2}}\,\int_{0}^{\infty }d\xi \,\,\alpha
_{33}(i\xi )\int_{\xi }^{\infty }d\gamma \,\frac{\gamma ^{3}}{\sqrt{\gamma
^{2}-\xi ^{2}}}\frac{K_{0}(a\gamma )}{I_{0}(a\gamma )}.  \label{Ub0}
\end{equation}%
For the evaluation of the CP force we need also the next-to-leading order
term near the string. For $q>2$, the dominant contribution comes from the $%
m=0$ term and the potential is given by the formula%
\begin{equation}
U_{\text{b}}(r)\approx U_{\text{b}}(0)-\frac{qr^{2}}{2\pi ^{2}}%
\,\int_{0}^{\infty }d\xi \,\,\alpha _{33}(i\xi )\int_{\xi }^{\infty }d\gamma
\,\frac{\gamma ^{5}}{\sqrt{\gamma ^{2}-\xi ^{2}}}\frac{K_{0}(a\gamma )}{%
I_{0}(a\gamma )}.  \label{Ub1a}
\end{equation}%
The corresponding CP force linearly vanishes on the string. In the case $%
1<q<2$, the dominant contribution to the next-to-leading order term comes
from the $m=1$ term and for the potential near the string one has%
\begin{eqnarray}
U_{\text{b}}(r) &\approx &U_{\text{b}}(0)-\frac{qr^{2q-2}}{2^{2q-1}\pi
^{2}\Gamma ^{2}(q)}\,\int_{0}^{\infty }d\xi \,\,\left[ \alpha _{11}(i\xi
)+\alpha _{22}(i\xi )\right]  \notag \\
&&\times \int_{\xi }^{\infty }d\gamma \,\frac{\gamma ^{2q-1}}{\sqrt{\gamma
^{2}-\xi ^{2}}}\,\,\left[ \frac{K_{q}(a\gamma )}{I_{q}(a\gamma )}\left(
\gamma ^{2}-\xi ^{2}\right) -\xi ^{2}\,\,\frac{K_{q}^{\prime }(a\gamma )}{%
I_{q}^{\prime }(a\gamma )}\right] .\,  \label{Ub1b}
\end{eqnarray}%
The corresponding CP force vanishes on the string for $q>1.5$ and diverges
for $q<1.5$. The boundary-free part in the potential near the string behaves
as $r^{-3}$ and it dominates.

For the isotropic polarizability tensor the general expression (\ref{Ubin})
for the boundary-induced part in the CP potential takes the form%
\begin{eqnarray}
U_{\text{b}}(r) &=&-\frac{2q}{\pi ^{2}}\,\sideset{}{'}{\sum}_{m=0}^{\infty
}\int_{0}^{\infty }d\xi \,\,\alpha (i\xi )\int_{\xi }^{\infty }\frac{\gamma
d\gamma }{\sqrt{\gamma ^{2}-\xi ^{2}}}  \notag \\
&&\,\times \left\{ \frac{K_{qm}(a\gamma )}{I_{qm}(a\gamma )}\left[ \left(
\gamma ^{2}-\xi ^{2}\right) F_{qm}(\gamma r)+\gamma ^{2}I_{qm}^{2}(\gamma r)%
\right] -\xi ^{2}\frac{K_{qm}^{\prime }(a\gamma )}{I_{qm}^{\prime }(a\gamma )%
}F_{qm}(\gamma r)\right\} .  \label{UbinIz}
\end{eqnarray}%
with the notation $F_{qm}(x)=I_{qm}^{\prime 2}(x)+(qm/x)^{2}I_{qm}^{2}(x)$.

The CP potential diverges on the cylindrical shell. For points near the
shell the dominant contribution comes from large values of $m$ and we can
use the uniform asymptotic expansions for the modified Bessel functions \cite%
{Abra72}. For the isotropic case, from (\ref{UbinIz}), to the leading order
one finds:%
\begin{equation}
U_{\text{b}}(r)\approx -\frac{(r-a)^{-4}}{16\pi }\,\int_{0}^{\infty }d\zeta
\,\alpha (i\zeta /[2(r-a)])e^{-\zeta }(\zeta ^{2}+2\zeta +2).  \label{Ubnear}
\end{equation}%
The expression in the right-hand side coincides with the CP potential for a
metallic plate in Minkowski spacetime, with $r-a$ being the distance from
the plate.

For the further transformation of the CP potential the polarizability tensor
should be specified. We use the anisotropic oscillator model:%
\begin{equation}
\alpha _{n}(i\xi )=\sum_{j}\frac{g_{j}^{(n)}}{\omega _{j}^{(n)2}+\xi ^{2}},
\label{alfam}
\end{equation}%
where $\omega _{j}^{(n)}$ and $g_{j}^{(n)}$ are the oscillator frequencies
and strengths, respectively. With this model, performing the integration in (%
\ref{Ubin}) we find%
\begin{eqnarray}
U_{\text{b}}(r) &=&-\frac{q}{\pi }\,\sideset{}{'}{\sum}_{m=0}^{\infty
}\sum_{n=1}^{3}\sum_{j}g_{j}^{(n)}\sum_{\lambda =0,1}\int_{0}^{\infty
}d\gamma \,\gamma \frac{K_{qm}^{(\lambda )}(a\gamma )}{I_{qm}^{(\lambda
)}(a\gamma )}  \notag \\
&&\times \lbrack \sqrt{1+\gamma ^{2}/\omega _{j}^{(n)2}}-1]f_{\lambda
,qm}(\gamma r,\sqrt{1+\gamma ^{2}/\omega _{j}^{(n)2}}),  \label{Ubin3}
\end{eqnarray}%
where we have introduced the notations%
\begin{eqnarray}
f_{0,qm}(x,y) &=&\beta _{1n}^{2}I_{qm}^{\prime 2}(x)+\beta _{2n}^{2}\left(
\frac{qm}{x}\right) ^{2}I_{qm}^{2}(x)+\left( 1+\frac{1}{y}\right) \beta
_{3n}^{2}I_{qm}^{2}(x),  \notag \\
f_{1,qm}(x,y) &=&-\frac{1}{y}\left[ \beta _{1n}^{2}\left( \frac{qm}{x}%
\right) ^{2}I_{qm}^{2}(x)+\beta _{2n}^{2}I_{qm}^{\prime 2}(x)\right] .
\label{flam}
\end{eqnarray}%
The coefficients $\beta _{ln}$ depend on the orientation of the
polarizability tensor principal axes with respect to the string.

In the isotropic case%
\begin{eqnarray}
U_{\text{b}}(r) &=&\frac{q}{\pi }\,\sideset{}{'}{\sum}_{m=0}^{\infty
}\sum_{j}\frac{g_{j}}{\omega _{j}^{2}}\int_{0}^{\infty }d\gamma \,\gamma
^{3}\left\{ \frac{K_{qm}^{\prime }(a\gamma )}{I_{qm}^{\prime }(a\gamma )}\,%
\frac{F_{qm}(\gamma r)}{s_{j}(\gamma )[s_{j}(\gamma )+1]}\right.  \notag \\
&&\left. -\frac{K_{qm}(a\gamma )}{I_{qm}(a\gamma )}\left[ \,\,\frac{%
F_{qm}(\gamma r)}{s_{j}(\gamma )+1}+\frac{I_{qm}^{2}(\gamma r)}{s_{j}(\gamma
)}\right] \right\} ,  \label{Ubiniz3}
\end{eqnarray}%
with the notation%
\begin{equation}
s_{j}(\gamma )=\sqrt{1+\gamma ^{2}/\omega _{j}^{2}}.  \label{sj}
\end{equation}%
For the boundary-induced part in the CP force we have $\mathbf{F}_{\text{b}%
}=F_{\text{b},r}\mathbf{n}_{r}$, where $\mathbf{n}_{r}$ is the unit vector
along the radial coordinate $r$ and $F_{\text{b},r}=-\partial _{r}U_{\text{b}%
}(r)$. Now, by using the inequality $I_{qm}^{\prime 2}(x)\leqslant \lbrack
1+(qm/x)^{2}]I_{qm}^{2}(x)$, from (\ref{Ubiniz3}) it can be seen that $%
\partial _{r}U_{\text{b}}(r)<0$. Hence, in the isotropic case one has $F_{%
\text{b},r}>0$ and the boundary-induced part in the CP force inside the
cylindrical shell is directed toward the shell. The pure string part of the
force has the same direction and the total force in the isotropic case is
repulsive with respect to the string and attractive with respect to the
shell.

\section{Green tensor and the Casimir-Polder potential in the exterior region%
}

\label{sec:GreenCPext}

In this section we consider the region outside the cylindrical boundary. The
corresponding mode functions for the electric field are given by formulas (%
\ref{Ealf}) and (\ref{Eir}) with the replacement (see \cite{Beze07})%
\begin{equation}
J_{q|m|}(\gamma r)\rightarrow g_{q|m|}^{(\lambda )}(\gamma a,\gamma
r)=J_{q|m|}(\gamma r)Y_{q|m|}^{(\lambda )}(\gamma a)-Y_{q|m|}(\gamma
r)J_{q|m|}^{(\lambda )}(\gamma a),  \label{extreplace}
\end{equation}%
and with the normalization coefficient%
\begin{equation}
\beta _{\alpha }^{-2}=\frac{2\pi }{q}\gamma \omega \left[ J_{q|m|}^{(\lambda
)2}(\gamma a)+Y_{q|m|}^{(\lambda )2}(\gamma a)\right] .  \label{betalfext}
\end{equation}%
Here, as before, $\lambda =0,1$ correspond to the waves of the electric and
magnetic types, respectively. Substituting the eigenfunctions into the
corresponding mode-sum formula, for the retarded Green tensor we find:%
\begin{eqnarray}
&&G_{jl}(x,x^{\prime })=-i\frac{q\theta (\tau )}{2\pi }\sum_{m=-\infty
}^{+\infty }\sum_{\lambda =0,1}\int_{-\infty }^{+\infty }dk\int_{0}^{\infty
}d\gamma \frac{(\gamma \omega )^{-1}}{J_{q|m|}^{(\lambda )2}(\gamma
a)+Y_{q|m|}^{(\lambda )2}(\gamma a)}  \notag \\
&&\qquad \times \left[ e^{iqm\Delta \phi +ik\Delta z-i\omega \Delta t}\bar{E}%
_{j}^{(\lambda )}(r)\bar{E}_{l}^{(\lambda )\ast }(r^{\prime })-e^{-iqm\Delta
\phi -ik\Delta z+i\omega \Delta t}\bar{E}_{l}^{(\lambda )}(r^{\prime })\bar{E%
}_{j}^{(\lambda )\ast }(r)\right] ,  \label{Gjlext}
\end{eqnarray}%
where the expressions for the functions $\bar{E}_{l}^{(\lambda )}(r)$ are
given by (\ref{Eir}) with the replacement (\ref{extreplace}).

For the further transformation of the expression for the Green tensor, we
use the identity%
\begin{equation}
\frac{\bar{E}_{j}^{(\lambda )}(r)\bar{E}_{l}^{(\lambda )\ast }(r^{\prime })}{%
J_{q|m|}^{(\lambda )2}(\gamma a)+Y_{q|m|}^{(\lambda )2}(\gamma a)}%
=E_{j}^{(\lambda )}(r)E_{l}^{(\lambda )\ast }(r^{\prime })-\frac{1}{2}%
\sum_{s=1}^{2}\frac{J_{q|m|}^{(\lambda )}(\gamma a)}{H_{q|m|}^{(s,\lambda
)}(\gamma a)}E_{(s)j}^{(\lambda )}(r)\tilde{E}_{(s)l}^{(\lambda )}(r^{\prime
}),  \label{Identity}
\end{equation}%
where the expression for the functions $E_{(s)l}^{(\lambda )}(r)$ and $%
\tilde{E}_{(s)l}^{(\lambda )}(r)$ are obtained from the expressions for $%
E_{l}^{(\lambda )}(r)$ and $E_{l}^{(\lambda )\ast }(r)$ from (\ref{Eir}),
respectively, by the replacement $J_{q|m|}(\gamma r)\rightarrow
H_{q|m|}^{(s)}(\gamma r)$. The part of the Green tensor corresponding to the
first term in the right-hand side of (\ref{Identity}) coincides with the
Green tensor in the boundary-free geometry, $G_{jl}^{(0)}(x,x^{\prime })$.
As a result, the Green tensor is decomposed as%
\begin{equation}
G_{jl}(x,x^{\prime })=G_{jl}^{(0)}(x,x^{\prime })+G_{jl}^{\text{(b)}%
}(x,x^{\prime }),  \label{GjlExtdec}
\end{equation}%
where the expression for the shell-induced part $G_{jl}^{\text{(b)}%
}(x,x^{\prime })$ is directly obtained from (\ref{Gjlext}) and (\ref%
{Identity}). For the corresponding spectral component we find%
\begin{eqnarray}
G_{jl}^{\text{(b)}}(\mathbf{r},\mathbf{r}^{\prime };i\xi ) &=&\frac{q}{4\pi }%
\sum_{m=-\infty }^{+\infty }\sum_{\lambda =0,1}\int_{-\infty }^{+\infty
}dk\sum_{s=1}^{2}\int_{0}^{\infty }d\gamma \frac{1}{\gamma \omega }\frac{%
J_{q|m|}^{(\lambda )}(\gamma a)}{H_{q|m|}^{(s)(\lambda )}(\gamma a)}\,
\notag \\
&&\times \left[ E_{(s)j}^{(\lambda )}(r)\tilde{E}_{(s)l}^{(\lambda
)}(r^{\prime })\frac{e^{iqm\Delta \phi +ik\Delta z}}{\omega -i\xi }%
+E_{(s)l}^{(\lambda )}(r^{\prime })\tilde{E}_{(s)j}^{(\lambda )}(r)\frac{%
e^{-iqm\Delta \phi -ik\Delta z}}{\omega +i\xi }\right] .  \label{GjlbExt2}
\end{eqnarray}%
For the term with $s=1$ ($s=2$) we rotate the contour of the integration
over $\gamma $ by $\pi /2$ ($-\pi /2$). After introducing the modified
Bessel functions, this leads to the final expression
\begin{eqnarray}
G_{jl}^{\text{(b)}}(\mathbf{r},\mathbf{r}^{\prime };i\xi ) &=&-\frac{q}{\pi }%
\,\sum_{m=-\infty }^{\infty }e^{iqm\Delta \phi }\sum_{\lambda =0,1}(-\xi
^{2})^{\lambda }\int_{-\infty }^{\infty }dk\,e^{ik\Delta z}  \notag \\
&&\times k^{2(1-\lambda )}\frac{I_{q|m|}^{(\lambda )}(a\gamma )}{%
K_{q|m|}^{(\lambda )}(a\gamma )}e_{j}^{(\lambda )}(\gamma r,\gamma
/k)e_{l}^{(\lambda )\ast }(\gamma r^{\prime },\gamma /k),  \label{GjlbExt3}
\end{eqnarray}%
where $\gamma =\sqrt{k^{2}+\xi ^{2}}$ and we have defined the functions
\begin{eqnarray}
e_{1}^{(0)}(x,y) &=&K_{q|m|}^{\prime }(x),\;e_{2}^{(0)}(x,y)=i\frac{qm}{x}%
K_{q|m|}(x),\;e_{3}^{(0)}(x,y)=iyK_{q|m|}(x),  \notag \\
e_{1}^{(1)}(x,y) &=&\frac{qm}{x}K_{q|m|}(x),\;e_{2}^{(1)}(x,y)=-iK_{q|m|}^{%
\prime }(x),\;e_{3}^{(1)}(x,y)=0.  \label{el}
\end{eqnarray}%
In the coincidence limit the off-diagonal components of the Green tensor
vanish.

By taking into account the expression (\ref{GjlbExt3}) of the Green tensor,
for the CP potential outside a cylindrical shell we get%
\begin{eqnarray}
U_{\text{b}}(r) &=&-\frac{2q}{\pi ^{2}}\,\sideset{}{'}{\sum}_{m=0}^{\infty
}\sum_{\lambda =0,1}\sum_{l=1}^{3}\int_{0}^{\infty }d\xi \,\,\alpha
_{ll}(i\xi )(-\xi ^{2})^{\lambda }  \notag \\
&&\times \,\int_{\xi }^{\infty }d\gamma \,\gamma \frac{I_{qm}^{(\lambda
)}(a\gamma )}{K_{qm}^{(\lambda )}(a\gamma )}\frac{|e_{l}^{(\lambda )}(\gamma
r,\gamma /\sqrt{\gamma ^{2}-\xi ^{2}})|^{2}}{\left( \gamma ^{2}-\xi
^{2}\right) ^{\lambda -1/2}}.  \label{Ubext}
\end{eqnarray}%
In the special case $q=1$, from this formula we obtain the result of \cite%
{Eber09,Beze11b} for the interaction potential with a cylindrical shell in
Minkowski spacetime. For the isotropic polarizability tensor the explicit
expression has the form%
\begin{eqnarray}
U_{\text{b}}(r) &=&-\frac{2q}{\pi ^{2}}\,\sideset{}{'}{\sum}_{m=0}^{\infty
}\int_{0}^{\infty }d\xi \,\,\alpha (i\xi )\int_{\xi }^{\infty }\frac{\gamma
d\gamma }{\sqrt{\gamma ^{2}-\xi ^{2}}}  \notag \\
&&\,\times \left\{ \frac{I_{qm}(a\gamma )}{K_{qm}(a\gamma )}\left[ \left(
\gamma ^{2}-\xi ^{2}\right) G_{qm}(\gamma r)+\gamma ^{2}K_{qm}^{2}(\gamma r)%
\right] -\xi ^{2}\frac{I_{qm}^{\prime }(a\gamma )}{K_{qm}^{\prime }(a\gamma )%
}G_{qm}(\gamma r)\right\} .  \label{Ubextiz}
\end{eqnarray}%
with the notation $G_{qm}(x)=K_{qm}^{\prime
2}(z)+(qm/z)^{2}K_{qm}^{2}(\gamma r)$. These formulas are obtained from the
corresponding formulas in the interior region by the interchange $%
I_{qm}\rightleftarrows K_{qm}$. Near the boundary the leading term in the
corresponding asymptotic expansion over the distance from the shell
coincides with (\ref{Ubnear}).

At large distances from the cylinder the dominant contribution comes from
the lower limit of the integration. Expanding the modified Bessel functions
for small values of the argument we can see that the dominant contribution
comes from the term $m=0$, $\lambda =0$ and to the leading order we find%
\begin{equation}
U_{\text{b}}(r)\approx -\frac{q\alpha _{11}(0)}{6\pi r^{4}\ln (r/a)}\,.
\label{Ubfar}
\end{equation}%
Note that at large distances the leading term in the pure string part is
given by the expression
\begin{equation}
U_{0}(r)\approx \frac{\left( q^{2}-1\right) \left( q^{2}+11\right) }{360\pi
r^{4}}\left[ \alpha _{11}(0)-\alpha _{22}(0)+\alpha _{33}(0)\right] .
\label{U1Large1}
\end{equation}%
Hence, at large distances the Casimir-Polder potential is
dominated by the pure string part and the corresponding force is
repulsive. As it follows from (\ref{Ubfar}), at large distances
the relative contribution of the boundary-induced effects in the
CP potential decays logarithmically. Considering the cylindrical
boundary as a simple model for string's core, we see that the
internal structure of the string may have non-negligible effects
even at large distances (see also the discussion in
\cite{Alle90}).

In the oscillator model for the polarizability tensor the expression for the
CP potential in the exterior region is obtained from (\ref{Ubin3}) by the
replacements $I_{qm}\rightleftarrows K_{qm}$. In particular, in the
isotropic case we have%
\begin{eqnarray}
U_{\text{b}}(r) &=&\frac{q}{\pi }\,\sideset{}{'}{\sum}_{m=0}^{\infty
}\sum_{j}\frac{g_{j}}{\omega _{j}^{2}}\int_{0}^{\infty }d\gamma \,\gamma
^{3}\left\{ \frac{I_{qm}^{\prime }(a\gamma )}{K_{qm}^{\prime }(a\gamma )}\,%
\frac{G_{qm}(\gamma r)}{s_{j}(\gamma )[s_{j}(\gamma )+1]}\right.  \notag \\
&&\left. -\frac{I_{qm}(a\gamma )}{K_{qm}(a\gamma )}\left[ \,\,\frac{%
G_{qm}(\gamma r)}{s_{j}(\gamma )+1}+\frac{K_{qm}^{2}(\gamma r)}{s_{j}(\gamma
)}\right] \right\} ,  \label{Ubext4}
\end{eqnarray}%
with $s_{j}(\gamma )$ defined by the relation (\ref{sj}). By using the
inequality $K_{qm}^{\prime 2}(x)\geqslant \lbrack 1+(qm/x)^{2}]K_{qm}^{2}(x)$%
, from (\ref{Ubext4}) it can be seen that in the exterior region $\partial
_{r}U_{\text{b}}(r)>0$. Consequently, the radial component of the
boundary-induced part in the CP force is negative, $F_{\text{b},r}=-\partial
_{r}U_{\text{b}}(r)<0$, and this force is attractive with respect to the
cylinder. For the isotropic case the radial component of the pure string
part in the force is positive and it has an opposite direction with respect
to the boundary-induced part. Near the cylindrical shell the
boundary-induced part dominates and the total force in the exterior region
is directed toward the cylinder. At large distances from the shell the pure
string part is dominant and the total force is repulsive with respect to the
cylinder.

\begin{figure}[tbph]
\begin{center}
\begin{tabular}{cc}
\epsfig{figure=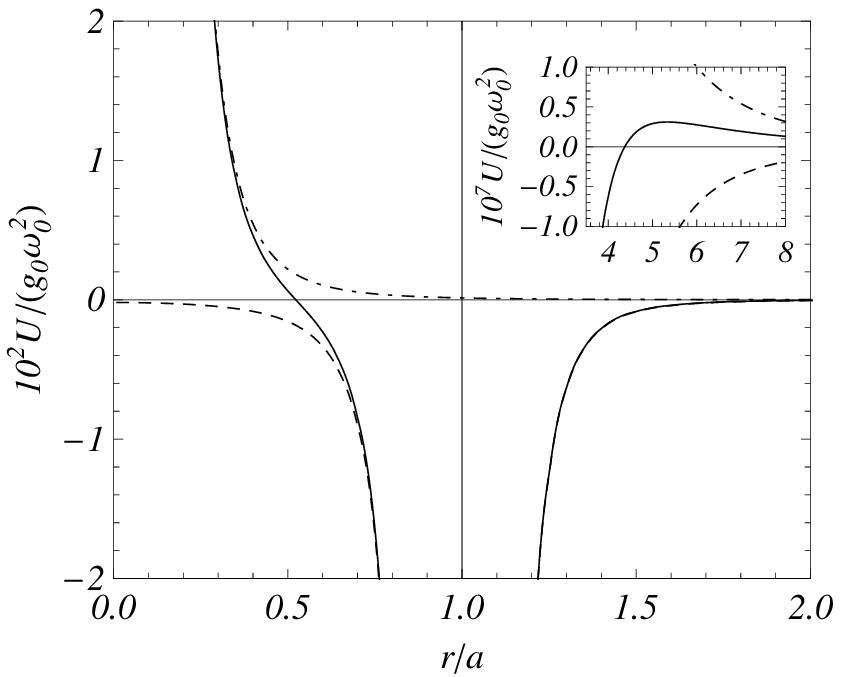,width=7.cm,height=6.cm} & \quad %
\epsfig{figure=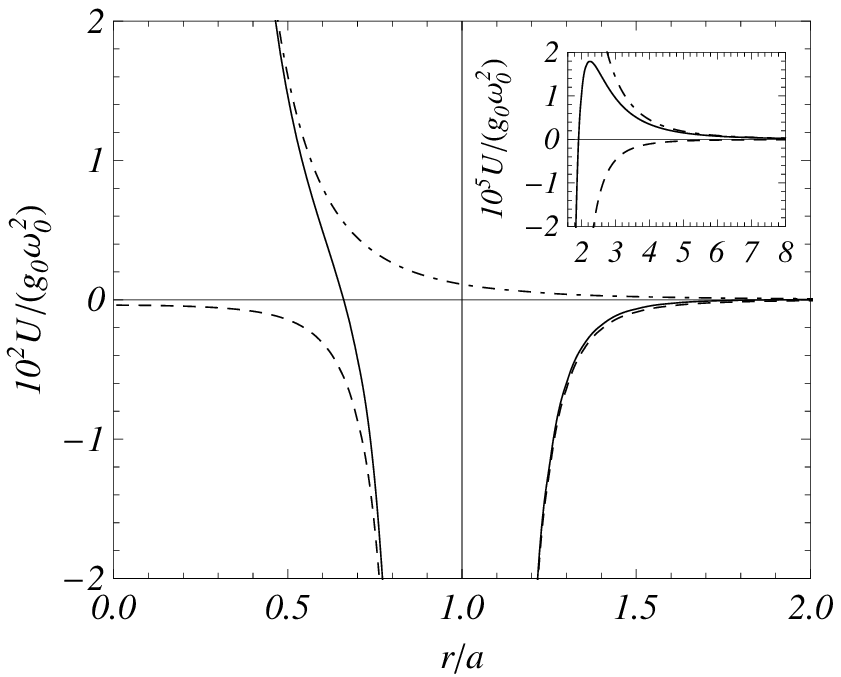,width=7.cm,height=6.cm}%
\end{tabular}%
\end{center}
\caption{Total CP potential (full curve), the pure string part (dot-dashed
curve) and the boundary-induced part (dashed curve) as functions of $r/a$
for $a/\protect\lambda _{0}=1$. The left and right panels are plotted for $%
q=2$ and $q=4$, respectively.}
\label{fig1}
\end{figure}

In figure \ref{fig1} we display the total CP potential $U(r)$ (full curve),
pure string part $U_{0}(r)$ (dot-dashed curve), and the boundary-induced
part $U_{\text{b}}(r)$ (dashed curve) as functions of the ratio $r/a$ for $%
a/\lambda _{0}=1$, with $\lambda _{0}=2\pi /\omega _{0}$. The single
oscillator model is used with isotropic polarizability and with the
parameters $g_{j}=g_{0}$ and $\omega _{j}=\omega _{0}$. The left and right
panels correspond to $q=2$ and $q=4$, respectively. As it has been explained
before, the potential is dominated by boundary-induced part near the
cylindrical shell and by the pure string part for points near the string and
at large distances from the cylindrical shell.

In the discussion above we have considered the idealized geometry with a
zero thickness cosmic string. A realistic cosmic string has a structure on a
length scale defined by the phase transition at which it is formed. In the
presence of a conducting cylindrical boundary, the CP potential in the
exterior region is uniquely defined by the boundary conditions and the bulk
geometry. From here it follows that if we consider a non-trivial core model
with finite thickness $b<a$ and with the line element (\ref{ds21}) in the
region $r>b$, the results in the region outside the cylindrical shell will
not be changed.

\section{Conclusion}

\label{sec:Conc}

We have investigated the CP potential for a polarizable microparticle in the
geometry of a straight cosmic string with a coaxial conducting cylindrical
shell. Both regions inside and outside the shell are considered. We start
the consideration from the evaluation of the Green tensor inside the shell.
In this region the mode sum contains the summation over the corresponding
eigenmodes which are expressed in terms of the zeros of the Bessel function
and its derivative. For the summation of the series over these zeros we have
employed the formula (\ref{Sumf}). This allowed to extract from the Green
tensor the part corresponding to the cosmic string geometry without
boundaries. The latter was previously investigated in \cite{Saha11EPJ} and
here we are mainly interested in the boundary-induced part. This part is
presented in the form (\ref{Gjl3}). For the evaluation of the CP potential
we need the Green tensor in the coincidence limit of the arguments. In this
limit the off-diagonal components vanish and the diagonal components are
given by (\ref{Gll}).

Similar to the case of the Green tensor, the CP potential is decomposed into
the pure string and boundary-induced parts. In the region inside the shell,
the latter is given by the expression (\ref{Ubin}). The CP potential depends
on the distance from the string and on the angles determining the
orientation of the principal axes of the polarizability tensor with respect
to the cosmic string. For the isotropic polarizability the general
expression is simplified to (\ref{UbinIz}). Unlike to the pure string part,
the boundary-induced part in the CP potential is finite on the string. The
corresponding asymptotic behavior near the string is given by expressions (%
\ref{Ub1a}) and (\ref{Ub1b}) for $q>2$ and $1<q<2$, respectively. The
boundary-induced part in the CP potential diverges on the cylindrical shell.
The leading term in the asymptotic expansion over the distance from the
shell coincides with the CP potential for a metallic plate in Minkowski
spacetime. As a model for a polarizability tensor we have used the
anisotropic oscillator model. The expressions for the CP potential with this
model are given by (\ref{Ubin3}) and (\ref{Ubiniz3}) for anisotropic and
isotropic cases respectively. In the isotropic case, the boundary-induced
part in the CP force inside the cylindrical shell is directed toward the
shell. The pure string part of the force has the same direction and the
total force in the isotropic case is repulsive with respect to the string
and attractive with respect to the shell.

The electromagnetic field Green tensor and the CP potential outside a
cylindrical shell have been discussed in section \ref{sec:GreenCPext}. By
making use of the identity (\ref{Identity}), the Green tensor is presented
as the sum of boundary-free and boundary-induced parts. The latter is given
by the expression (\ref{GjlbExt3}). The corresponding expressions for the CP
potential have the form (\ref{Ubext}) and (\ref{Ubextiz}) for the
anisotropic and isotropic polarizabilities, respectively. At large distances
from the cylinder the leading term in the boundary-induced CP potential is
given by the expression (\ref{Ubfar}). The leading term in the pure string
part is given by the expression (\ref{U1Large1}) and it dominates at large
distances. The corresponding force is repulsive. For the oscillator model,
the expression for the boundary-induced CP potential takes the form (\ref%
{Ubext4}). In the isotropic case the corresponding force is attractive with
respect to the cylinder. For the isotropic case the pure string part in the
force has an opposite direction with respect to the boundary-induced part.
Near the cylindrical shell the boundary-induced part dominates and the total
force in the exterior region is directed toward the cylinder. At large
distances from the shell the pure string part is dominant and the total
force is repulsive with respect to the cylinder. From the point of view of
the physics in the exterior region the conducting cylindrical surface can be
considered as a simple model of superconducting string core. Superconducting
strings are predicted in a wide class of field theories and they are sources
of a number of interesting astrophysical effects such as generation of
synchrotron radiation, cosmic rays, and relativistic jets.

\section*{Acknowledgments}

A.A.S. was partially supported by PVE/CAPES Program (Brazil). A.A.S.
gratefully acknowledges the hospitality of the Federal University of Para%
\'{\i}ba (Jo\~{a}o Pessoa, Brazil) where part of this work was done.

\end{document}